\documentclass[iop]{emulateapj}
\accepted{ to AJ {September 16, 2013}}
\usepackage{natbib}
\usepackage{graphicx}
\usepackage{mathtools}
\usepackage{multirow}

\newcommand{\cothree}{\mbox{\rm CO(3\,--\,2)}}
\newcommand{\cotwo}{\mbox{\rm CO(2\,--\,1)}}
\newcommand{\coone}{\mbox{\rm CO(1\,--\,0)}}
\newcommand{\hi}{\mbox{\rm H$\,$\scshape{i}}}
\newcommand{\hii}{\mbox{\rm H$\,$\scshape{ii}}}
\newcommand{\kmpers}{\mbox{km~s$^{-1}$}}
\newcommand{\Kkmpers}{\mbox{K~km~s$^{-1}$}}

\newcommand{\xco}{\mbox{$X_{\rm CO}$}}

\newcommand{\MJypersr}{\mbox{MJy~sr$^{-1}$}}
\newcommand{\Msunperpc}{\mbox{\rm M$_{\odot}$ pc$^{-2}$}}

\newcommand{\Msunperyrperkpc}{\mbox{\rm M$_{\odot}$ yr$^{-1}$ kpc$^{-2}$}}

\newcommand{\hisigmean}{$12.7 \pm 3.1$ \kmpers}
\newcommand{\hivmean}{$12.7$ \kmpers}
\newcommand{\hidisp}{$3.1$ \kmpers}
\newcommand{\cosigmean}{$12.8 \pm 3.9$ \kmpers}
\newcommand{\covmean}{$12.8$ \kmpers}
\newcommand{\counc}{$3.9$ \kmpers}
\newcommand{\ratmean}{$1.0 \pm 0.2$}

\newcommand{\hisigmed}{$11.9 \pm 3.1$ \kmpers}
\newcommand{\hivmed}{$11.9$ \kmpers}
\newcommand{\cosigmed}{$12.0 \pm 3.9$ \kmpers}

\shorttitle{High--Dispersion Molecular Gas}
\shortauthors{Cald\'{u}-Primo et al.}
\slugcomment{Draft}

\begin{document}

\title{A High--Dispersion Molecular Gas Component in Nearby Galaxies}

\author{
Anahi Cald\'{u}-Primo\altaffilmark{1},
Andreas Schruba\altaffilmark{2},
Fabian Walter\altaffilmark{1},
Adam Leroy\altaffilmark{3},
Karin Sandstrom\altaffilmark{1}, \linebreak 
W.J.G. de Blok\altaffilmark{4,5}, 
R.Ianjamasimanana\altaffilmark{4},
K.M. Mogotsi\altaffilmark{4},
}

\altaffiltext{1}{Max-Planck-Institut f\"ur Astronomie, K\"onigstuhl 17, 69117 Heidelberg, Germany; caldu@mpia.de}
\altaffiltext{2}{Cahill Center for Astronomy and Astrophysics, California Institute of Technology, MC 249-17, 1200 E California Blvd, Pasadena, CA 91125, USA}
\altaffiltext{3}{National Radio Astronomy Observatory, 520 Edgemont Road Charlottesville, VA 22903, USA}
\altaffiltext{4}{Astrophysics, Cosmology and Gravity Centre (ACGC), Department of Astronomy, University of Cape Town, Private Bag X3, Rondebosch 7701, South Africa}
\altaffiltext{5}{Netherlands Institute for Radio Astronomy (ASTRON), Postbus 2, 7990 AA Dwingeloo, Netherlands}

\begin{abstract}
We present a comprehensive study of the velocity dispersion of the atomic (\hi) and molecular (H$_2$) gas components in the disks (R $\lesssim\ R_{25}$) of a sample of 12 nearby spiral galaxies with moderate inclinations. Our analysis is based on sensitive high resolution data from the THINGS (atomic gas) and HERACLES (molecular gas) surveys. To obtain reliable measurements of the velocity dispersion, we stack  regions several kilo-parsecs in size, after accounting for intrinsic velocity shifts due to galactic rotation and large--scale motions. We stack using various parameters: the galacto-centric distance, star formation rate surface density, \hi\ surface density, H$_2$ surface density, and total gas surface density. We fit single Gaussian components to the stacked spectra and measure median velocity dispersions for \hi\ of \hisigmed\ and for H$_2$ of \cosigmed. The CO velocity dispersions are thus, surprisingly, very similar to the corresponding ones of \hi, with an average ratio of $\sigma_{\rm HI}/\sigma_{\rm CO}=$ \ratmean\ irrespective of  the stacking parameter. The measured CO velocity dispersions are significantly higher (factor $\sim$2) than the traditional picture of a cold molecular gas disk associated with star formation. The high dispersion implies an additional thick molecular gas disk (possibly as thick as the \hi\ disk). Our finding is in agreement with recent sensitive measurements in individual edge--on and face--on galaxies and points towards the general existence of a thick disk of molecular gas, in addition to the well--known thin disk in nearby spiral galaxies.
\end{abstract}


\section{Introduction}
\label{intro}
The interstellar medium (ISM) gas in a galaxy exists in different phases, as a result of complex radiative, thermal, and kinematic processes, that are present in complex galactic structures. These phases are continuously exchanging energy among each other, they are heated by stellar radiation and they cool by line and continuum emission. While the physical processes which heat, cool, and transport energy in the ISM are generally known, predicting their detailed balance via theoretical models or simulations is complicated by the vast number of processes and dynamic range of each component which alter the rate of each reaction. Therefore, sensitive observations are indispensable to resolve the structure of the different ISM phases and shed light on their balance.

The classical picture of the ISM classifies gas in five phases \citep{mi81}: hot and warm ionized phases (HIM, WIM), warm and cold neutral atomic phases (WNM, CNM), and a molecular phase (H$_2$). Typical properties of these phases are thought to be: The warm ionized medium has T\,$\sim$\,10$^{4}$\,K, n\,$\sim$\,0.3\,cm$^{-3}$, and a local volume filling factor f\,$\geq$\,15\%. Although it is mainly associated with \hii\, regions, observations have shown that a considerable fraction of the ISM and extraplanar gas is filled with ionized gas \citep{re84}. The hot ionized medium with T\,$\sim$\,10$^{6}$\,K, n\,$\sim$\,10$^{-3}$\,cm$^{-3}$, and f $\leq$ 50 \% is believed to be produced in supernovae explosions.

The neutral atomic gas is predicted to separate into two phases: a clumpy cold neutral medium (CNM) embedded in a more diffuse  warm neutral medium (WNM) \citep{fi69,wo95,wo03}.  These two phases are thought to coexist in pressure equilibrium. The CNM has T\,$\gtrsim$\,300\,K, n\,$\sim$\,20\,cm$^{-3}$, and f\,$\sim$\,2--4\%, and is distributed in rather dense filaments, making up for a small fraction of the ISM. The WNM has temperatures just below 10$^{4}$\,K, n\,$\sim$\,0.1-0.3\,cm$^{-3}$, and f\,$\geq$\,30\%. Molecular gas has so far been predominantly found in giant molecular clouds (GMCs), mostly distributed near the midplane of a galaxy. The temperatures inside these clouds are of the order of 10\,K, densities of n\,$\geq$\,10$^{3}$\,cm$^{-3}$, and f\,$<$\,1\%.

These different phases do not have sharp boundaries. At least 50\% of the WNM may be in thermally unstable regions \citep{he03}, while molecular gas is presumably formed from the CNM \citep{el93,va06}. The temperature and density structures vary smoothly throughout the gas.  The CNM has characteristic thermal velocity dispersions of $\lesssim$\,1.5\,\kmpers, while the WNM has thermal velocity dispersions of $\sim$\,8\,\kmpers\,\citep{wo03}.

However, early studies \citep[][Garcia-Burillo et al., in prep]{ga92,co97,pe13}  have challenged this picture by pointing toward an additional diffuse low density molecular component with a scale height that is significantly higher than  the one that accounts for the majority of GMCs in a galaxy's disk. \citet{co97} studied the two nearly-face on galaxies NGC\,628 and NGC\,3938, and found similar \hi\ and CO vertical velocity dispersions, which, assuming an isothermal distribution, implies similar scale heights. Therefore, they conclude that both atomic and molecular gas make up a single dynamical component. The gas scale height of the edge-on galaxy NGC\,891 has been investigated by \citet[][and in prep.]{ga92}. They detect molecular emission at $\sim$1 kpc above the disk, which they also interpret as the presence of a low density thick disk component  (this is confirmed by new large--scale mapping of the galaxy with the IRAM 30\,m telescope, Garcia--Burillo, priv. comm.\footnote{We note that the interferometric study of NGC\,891 by \citet{sc93} did not recover this thick disk which is likely due to the fact that such a diffuse component is invisible to an interferometer.}). Very recently, \citet{pe13} published an in--depth study of M\,51 (also part of our sample) where they convincingly showed that M\,51 has a diffuse molecular disk with a velocity dispersion that is much larger than the molecular gas that is associated with GMCs in the star-forming disk.

So far most extragalactic studies concentrated exclusively on studying either \hi\, emission  \citep{ta09,wa12} or CO emission \citep{bo08}. In this paper we present a comprehensive study of the velocity dispersions of both the atomic and molecular gas phases within 12 nearby spiral galaxies. We use data from the VLA  THINGS \hi\ survey \citep{wa08} and from the IRAM HERACLES CO survey \citep{le09}. Even with the high sensitivity of instruments like the VLA or the IRAM 30-m telescope, stacking is useful to obtain average spectra of high significance and robust velocity dispersion measurements. The stacking technique we apply has been previously used by  \citet{sch11} to obtain highly sensitive measurements of \emph{integrated intensities} of the CO line out to large radii. In this paper we will use this method to determine robust measurements of \hi\, and CO \emph{velocity dispersions} within nearby galaxies.

The paper is structured as follows. In Section~\ref{data} we describe our galaxy sample and the \hi\ and CO data. In Section~\ref{method} we describe our stacking technique and method to determine velocity dispersions, including a detailed description of our uncertainty determinations. In Section~\ref{results} we give an interpretation and discuss implications of our results. Section~\ref{sum} summarizes our work.

\section{Data}
\label{data}

\subsection{Sample Selection}
We study 12 nearby star-forming spiral galaxies. These galaxies are drawn from a larger sample of galaxies targeted by various multi-wavelength surveys. The wavelengths we are interested in for the purpose of this study  are \hi\, (THINGS), CO (HERACLES), FUV \citep[GALEX NGS,][]{gi07}, and 24 $\mu$m \citep[SINGS,][]{ke03}. Our sample is chosen such that all galaxies are detected in these 4 data sets.  Given this restriction, we have a selection effect as we only study spiral galaxies with a clear detection in CO, our least sensitive tracer, in at least the inner parts of the galaxies. The THINGS survey also includes irregular galaxies, which we leave out from our analysis. During our analysis we convolve all data sets to the limiting $13^{\prime\prime}$ angular resolution of the HERACLES CO data. Galaxies for which this angular resolution corresponds to a linear resolution larger than 700\,pc (corresponding to a distance of $\sim$11 Mpc) are also left out. The last cut to obtain our working sample is based on a signal-to-noise cut of the stacked spectra (this will be explained in detail in Section~\ref{method}). Based on this we drop out the galaxies NGC 3521 and NGC 3627. The galaxies included in this work are listed in Table~\ref{tab1}, together with their adopted properties: distance, inclination, position angle, optical radius (R$_{25}$), and linear resolution (at $13^{\prime\prime}$ angular resolution), where the latter values are taken from \citet{wa08}. 

\begin{deluxetable}{c c c c c c} 
\tablecolumns{6}
\tablewidth{0pt}
\tablecaption{Properties of the galaxies used in this study \label{tab1}}

\tablehead{\colhead{Galaxy} & \colhead{D} & \colhead{Incl} & \colhead{PA} & \colhead{R$_{25}$} & \colhead{Res} \\
 & \colhead{Mpc} & \colhead{$^{\circ}$} & \colhead{$^{\circ}$} & \colhead{kpc} & \colhead{kpc} }
 \startdata
NGC \phn628 & \phn7.3 & \phn7 & \phn20 & 10.4 & 0.46 \\
NGC \phn925 & \phn9.2 & 66 & 287 & 14.2 & 0.58 \\
NGC 2403 & \phn3.2 & 63 & 124 & \phn7.3 & 0.20 \\
NGC 2903 & \phn8.9 & 65 & 204 & 15.3 & 0.56 \\
NGC 2976 & \phn3.6 & 65 & 335 & \phn3.8 & 0.23 \\
NGC 3184 & 11.1 & 16 & 179 & 11.9 & 0.70 \\
NGC 3351 & 10.1 & 41 & 192 & 10.6 & 0.64 \\
NGC 4736 & \phn4.7 & 41 & 296 & \phn5.3 & 0.30 \\
NGC 5055 & 10.1 & 59 & 102 & 17.4 & 0.64 \\
NGC 5194 & \phn8.0 & 20 & 172 & \phn9.0 & 0.50 \\
NGC 5457 & \phn7.4 & 18 & \phn39 & 25.8 & 0.47 \\
NGC 6946 & \phn5.9 & 33 & 243 & \phn9.8 & 0.37 
\enddata

\end{deluxetable}

\subsection{Atomic Gas Data}
The  \hi\ data come from the THINGS survey \citep{wa08}. This survey encompasses 34 nearby galaxies at distances of 2$-$15 Mpc. The observations have high spectral (2.6 or 5.2 \kmpers) and spatial ($\sim$11$^{\prime\prime}$ for natural weighting) resolutions. We use the so-called ``rescaled'' data cubes. This flux ``rescaling'' is important for computing correct fluxes, but also for obtaining accurate spectral shapes. More details can be found in \citet[Section 3.5 and references therein]{wa08}. Mass surface densities are derived from integrated line emission, i.e., zeroth moment, via
\begin{equation}
\Sigma_{\rm HI} = 0.02\ I_{\rm HI}\ \cos{\,i}
\end{equation}
\noindent where $\Sigma_{\rm HI}$ has units of \Msunperpc\ and includes a factor of $1.36$ to account for heavy elements, \textit{I}$_{\rm{HI}}$ is measured in \Kkmpers.

\subsection{Molecular Gas Data}
We obtain the \cotwo\ line emission from the cubes of the HERACLES survey \citep{le09}. The angular resolution of this survey is of $13^{\prime\prime}$ and it has a spectral resolution of $2.6$ \kmpers. The H$_{2}$ surface density is derived from the \cotwo line intensity by
\begin{equation}
\Sigma_{\rm H2} = 6.25\ I_{\rm CO}\ \cos{\,i}
\end{equation}
\noindent where $\Sigma_{\rm H2}$ has units of \Msunperpc\ again including heavy elements and \textit{I}$_{\rm CO}$ is measured in \Kkmpers. This conversion assumes a CO line ratio of $I_{\rm{CO}}(2-1)/I_{\rm{CO}}(1-0) = 0.7$ and a \coone -to-H$_{2}$  conversion factor \xco$ = 2.0 \times10^{20}$ cm$^{-2}$ \citep[see][and references therein]{le09}. We note that measurement of velocity dispersions are not affected by the choice of the conversion factor.

\subsection{Star Formation Tracer}
In order to determine the spatial distribution of recent star formation we use two tracers. The unobscured recent star formation is derived from the FUV emission. FUV traces O and B stars with typical ages of 20$-$30 Myr, and reaching sensitivities of up to $\sim$100 Myr \citep{sa07}. The embedded star formation can be accounted for using the mid-infrared 24 $\mu$m emission, which mainly comes from young stars' photons reprocessed by dust. The FUV images come from GALEX NGS \citep{gi07}. We make use only of the FUV band, which covers 1350$-$1750 \AA\ at an angular resolution of $\sim$4.5$^{\prime\prime}$. The photometry at 24 $\mu$m was obtained from the \textit{Spitzer} legacy survey SINGS \citep{ke03}. The images have a native resolution of $\sim$6$^{\prime\prime}$. Using these pair of tracers we determine the star formation rate surface density, $\Sigma_{\rm SFR}$, following \citet{le12}.
\begin{equation}
\Sigma_{\rm SFR}  = (8.1\times10^{-2}\,I_{\rm FUV} + 4.2 \times 10^{-3} I_{\rm 24\mu m}) \cos{\,i}
\end{equation}
\noindent where $\Sigma_{\rm SFR}$ has units of \Msunperyrperkpc, and FUV and 24 $\mu$m are measured in \MJypersr.

\section{Methodology}
\label{method}
 
\subsection{Stacking the Spectra}
Our aim is to obtain robust measurements of the velocity dispersions. This requires both high signal-to-noise (S/N) line profiles and a robust method to measure the velocity dispersion. To attain this requisite we average many spectra over significant sub-regions inside the galaxies. Individual spectra peak at different velocities in the galaxy due to rotation and large scale gas motions. \citet{sch11} used the first moment of the \hi\ emission to shift the spectra to a common velocity so that the shifted spectra could be coherently stacked. Following that work, we shift the spectra using the first \hi\ moment of intensity for NGC 628, NGC 925, NGC 3184, NGC 3351, NGC 4736, NGC 5194, NGC 5457, and NGC 6946. For NGC 2403, NGC 2903, NGC 2976, and NGC 5055 we have high--quality Hermite \emph{h$_{3}$} velocity maps \citep{bl08} available and we use those instead to perform the shifting. The latter velocity fields are superior to the first moment maps in constraining the `peak' of the emission \citep[see for details][]{bl08}, and indeed stacking yields velocity dispersions that are of the order of 1.5\,\kmpers\, lower than when using the first moment maps.

The procedure is then as follows: For each line of sight we take either the \hi\, or CO  spectrum and shift it by its Hermite \emph{h$_3$},  or its first moment map \hi\, velocity value. As shown in \citet{sch11}, the \hi\, mean velocity is an excellent proxy for the  CO mean velocity (we have also tested  this using the CO velocity field for the shifting, see discussion Section \ref{shhi}). Once all the spectra are shifted to a common velocity they are ready for coherent stacking over larger regions.

We stack the spectra based on different physical properties characterizing individual lines of sight: galactocentric radius, and star formation rate, atomic hydrogen, molecular hydrogen, and total gas surface densities. For each parameter we define bins that cover the complete range of parameter values measured in the galaxies and stack the shifted spectra inside these bins.

\subsubsection {Measurement of Velocity Dispersions}
After stacking the spectra we need a consistent way of measuring the velocity dispersion. For simplicity we will proceed to fit the data with a single Gaussian. Studies like \citet{pe07} show that even though a single-Gaussian function does not adequately describe the line shapes of \hi\, spectra, the width of the \hi\, line is well characterized by it. As the purpose of this study is to provide a global measurement of line widths, we fit each spectrum with a single Gaussian. To perform the fit we take a window centered on the peak of emission (in this case zero velocity after shifting) and of 100 \kmpers\ total width. We use the \texttt{MPFIT} function in \texttt{IDL}, which is a least-squares curve fitting procedure. The free parameters we determine by fitting are the line center velocity, the peak amplitude, and the velocity dispersion.

\begin{figure}[h]
\centering
\epsscale{1.1}\plotone{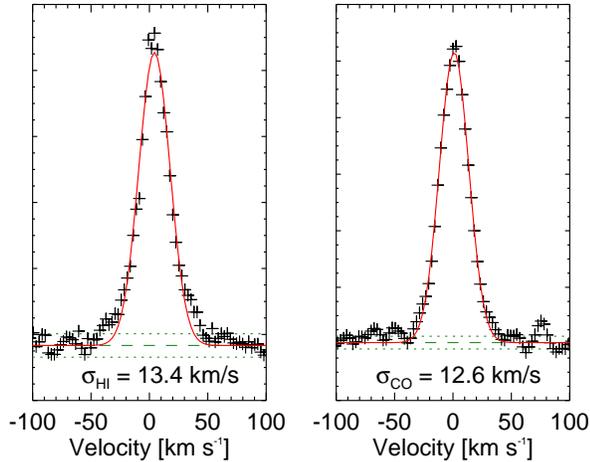}
\caption{Example of a stacked spectrum within NGC 6946 for a SFR bin of (8$-$10)$\times10^{-3}$ \Msunperyrperkpc. Left: \hi\, spectrum, Right: CO spectrum. The crosses are the measured values. We perform a Gaussian fit to the spectra (red solid line) in order to determine the velocity dispersion. The green dashed line represents the zero level, whereas the dotted green lines show the 1$\sigma$ rms noise level.}\label{fig:fig1}
\end{figure}

To ensure good S/N, we exclude (stacked) spectra with peak amplitudes lower than 5$\sigma$.  The fitting procedure allows us to check whether the spectra are best fitted with a double horn profile, which is usually the case for very broad spectra and spectra with clear multiple velocity components. After the fitting is done and to keep the analysis and determination of the velocity dispersion as simple and reliable as possible, we leave out double horn spectra which amount to 4.5\% of the high S/N spectra. These double-horn profiles are found in the center-most parts of galaxies (R $\lesssim$ 0.2 R$_{25}$), where generally strong non-Gaussian features are observed. At larger radii \hi\ spectra is in general smooth even beyond R$_{25}$. However, CO is more centrally concentrated  in spiral galaxies. \citet{sch11} studied the azimuthally averaged radial profiles of CO in a similar sample of galaxies and were able to detect CO out to $\sim$\,R$_{25}$. Here we restrict our analysis to the regions where we obtain high S/N emission in both tracers. In Figure~\ref{fig:fig1}  we show the resulting stacked \hi\, and $\rm CO$ spectra in NGC 6946 for one SFR bin as an example.

\subsubsection{Stacking by Galactocentric Radius}
Many galactic properties vary with radius. It is therefore interesting to investigate possible trends  in velocity dispersion as function of galactocentric radius. We use bins $15^{\prime\prime}$ wide, i.e., somewhat larger than our angular resolution. These bins have linear sizes that range from $260-800$ pc in our selected galaxies. We are investigating bins up to R$_{25}$ (for the equivalence in kpc see Table~\ref{tab1}) which gives $8-22$ bins per galaxy.

\subsubsection{Stacking by SFR Surface Density}
It is suggested that  the local rate of star formation is closely linked  to the velocity dispersion in the gas \citep[e.g.,][]{ta09}. To search for such a correlation we select regions of different levels of recent star formation rate density as traced by a combination of FUV and 24\,$\mu$m intensities. We define logarithmic bins of size of $0.1$ dex between $10^{-3}$ and $10^{-1}$ \Msunperyrperkpc. The number of bins per galaxy  vary between 8 and 20.

\subsubsection{Stacking by \hi\, Surface Density}
We also separate the lines of sight by different \hi\ mass surface densities. We define bins of 3 \Msunperpc\ within a range of $0-30$ \Msunperpc\ providing $3-9$ bins per galaxy.

\subsubsection{Stacking by H$_{2}$ Surface Density}
In this case we divide the lines of sight by H$_{2}$  mass surface density using bins 5 \Msunperpc\ wide and ranging over $0-50$ \Msunperpc.
The number of bins per galaxy  varies between 2 and 10.

\subsubsection{Stacking by Total Gas Surface Density}
For the total gas surface density, defined as $\Sigma_{\rm HI+H2} = \Sigma_{\rm HI} + \Sigma_{\rm H2}$, we again use bins 5 \Msunperpc\ wide covering a range of $0-50$ \Msunperpc\ giving us $3-10$ bins per galaxy.

\subsection{Uncertainties Determination}
\subsubsection{Shifting with \hi\ Velocity Fields}\label{shhi}
To align the \hi\ and CO spectral lines at a common velocity, we use the \hi\, mean velocity map or Hermite $h_3$ maps (where available) at its native resolution. This may induce an extra broadening in the stacked CO spectra in case there are significant deviations between the local CO velocity compared to the \hi\ velocity. \citet{sch11} tested for the difference between \hi\ and CO velocities for individual lines of sight at a $13^{\prime\prime}$ resolution (see their Fig.\,1).  They find a median $\overline{v}_{\rm CO} - \overline{v}_{\rm HI}$ of $-0.22$ \kmpers, i.e., consistent with no shift considering our channel width of $2.6$ \kmpers. In addition, we perform the following test on NGC 6946, one of the CO--brightest galaxies in our sample for which we can derive a reliable CO velocity field. Instead of using the \hi\ velocity field to shift the observed spectra before stacking, we use the respective CO velocity field. We carry out the subsequent analysis in the exact same way. The velocity dispersions of the stacked \hi\ and CO spectra have a median difference of less than 1~\kmpers, again smaller than the channel width. We conclude that no significant extra broadening is introduced to the CO stacked spectra by shifting using the \hi\ velocity fields. 

\subsubsection{Velocity Dispersions Obtained Without Stacking}
We also check whether shifting and stacking spectra naturally introduces a spurious widening in the measured velocity dispersions. We use NGC 628 and NGC 2403 to address this issue. Using the original data cube, we fit a Gaussian to each line of sight independently, i.e., without stacking any spectra. We exclude spectra using the same restrictions as with the stacked spectra (leaving out low signal to noise spectra and spectra with clear non-Gaussian features). Then for a given stacking region we compare the median value of velocity dispersions of individual lines of sight to the velocity dispersion of the stacked spectrum. Typically the latter values are larger by $1.4$ \kmpers\ with a 1$\sigma$ dispersion of $1.2$ \kmpers. Since this difference is again significantly smaller that our channel width of either $2.6$ or $5.2$ \kmpers, we deem this effect insignificant to our conclusions. A detailed analysis of fitting individual CO and \hi\ spectra in nearby galaxies are presented in Mogotsi et al. (in prep.).

\subsubsection{Galaxy Inclination}
\label{inclination}
\begin{figure}[h]
\centering
\epsscale{1.0}\plotone{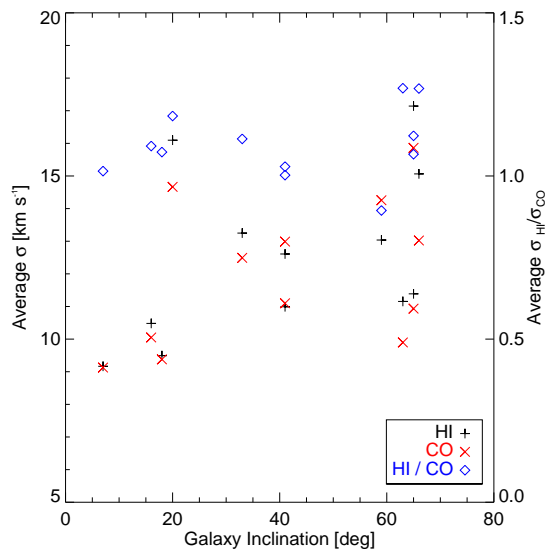}
\caption{Correlation of galaxy-averaged velocity dispersion with galaxy inclination for: {\em left y--axis:} atomic gas (black plus symbols) and molecular gas (red crosses), and {\em right y--axis:} their ratio (blue diamonds). The measured velocity dispersions correlate at most weakly with galaxy inclination.}
\label{fig:fig2}
\end{figure}

Galaxy inclination may affect the measured velocity dispersions as in highly inclined galaxies lines of sights probe larger lengths through the galactic disk. Thus one line of sight includes gas moving at different velocities leading to a broadening of the observed line profile. To test the importance of this effect, we show in Figure~\ref{fig:fig2}  for each galaxy  the average $\sigma_{\rm HI}$ value (black plus signs) and the average $\sigma_{\rm CO}$ value (red crosses) as function of galaxy inclination. In blue diamonds we show the ratio of $\sigma_{\rm HI}/\sigma_{\rm CO}$. We use the Spearman's rank correlation coefficients to quantify the magnitude of correlation. The coefficients we obtain are of 0.58, 0.48, and 0.27 for $\sigma_{\rm HI}$, $\sigma_{\rm CO}$, and $\sigma_{\rm HI}$/$\sigma_{\rm CO}$, respectively. These numbers tell us that there is not a significant correlation between velocity dispersion and galaxy inclination, as already apparent from Figure~\ref{fig:fig2}.

\subsubsection{Beam Smearing}
\label{galmod}

\begin{figure}[h]
\centering
\epsscale{1.1}\plotone{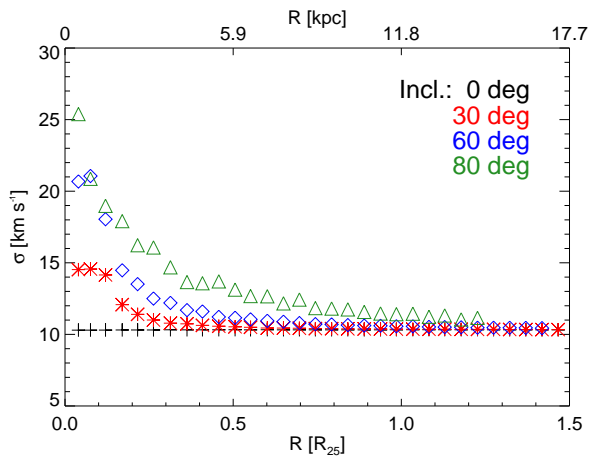}
\caption{Measured \hi\, velocity dispersion values obtained from model galaxies generated by the \texttt{GALMOD} task in \texttt{GIPSY}. We model a galaxy at various inclinations: black crosses for face-on, red asterisks for $30^{\circ}$ inclination, blue diamonds for $60^{\circ}$, and green triangles for $80^{\circ}$. Beam smearing becomes more important for highly inclined galaxies and in the centermost parts of galaxies. Thus, we exclude data at $R < 0.2\ R_{25}$ from the remaining analysis.}
\label{fig:fig3}
\end{figure}

As the telescope beam is not infinitely small, it inevitably averages over gas moving at different velocities. This will affect determined velocity values especially in regions of large velocity gradients, i.e., in the inner regions of galaxies, and in highly inclined galaxies. To quantify the beam smearing, we model galaxies using the task \texttt{GALMOD} in the Groningen Image Processing System (\texttt{GIPSY}\footnote{www.astro.rug.nl/\textasciitilde gipsy/}). This task allows us to model observations of the \hi\ gas in a spiral galaxy. We generate galaxies with $\Sigma_{\rm HI} = 5 \times 10^{21}$ cm$^{-2}$, velocity dispersion of $\sigma = 10$ \kmpers, and $sech^{2}$ vertical density profile, which are typical values for spiral galaxies.  All linear sizes were calculated assuming a distance of $7.8$ Mpc, the mean distance of our sample. We here present the results for a rotation curve which has a linear increase in velocity from 0 in the center of the galaxy reaching 200 \kmpers\ in the first 2~kpc ($52^{\prime\prime}$). This simple model rotation curve is in a broad sense the most similar to the average rotation curves of our sample. We use 4 different inclinations for the galaxies:  0, 30, 60, and 80 degrees and perform the exact same analysis as with the observed galaxies, accounting for resolution. We show the results on Figure~\ref{fig:fig3}. The optical radius value of $R_{25} = 11.8$ kpc is the average value of our sample of galaxies. Galaxies inclined by 80 degrees have 2.5 times larger `observed' velocity dispersions, especially toward the galaxy center. Galaxies at intermediate inclination (30 or 60 degrees) are only affected in the very central parts at $R \lesssim 0.2\ R_{25}$. On the other hand, for $R \gtrsim 0.2\ R_{25}$, the beam smearing effect is comparable or smaller than the channel width (2.6 or 5.2 \kmpers). Since the highest inclination of our galaxy sample is 66 degrees, by excluding points inside 0.2\ R$_{25}$ in the histogram plots (Figure \mbox{\ref{fig:fig5}}), we are left with points for which beam smearing only has a minor effect.

\subsubsection{Statistical Effects in Stacked Spectra}
\label{stat}
We want to test how the number of spectra and observational noise is affecting the resulting stacked spectra and derived velocity dispersion measurements. To test the former, we do the following: if for a certain bin the stacked spectrum consists of \emph{n} different individual spectra, we randomly draw \emph{n} times a spectrum from these spectra, allowing for repetition. We then stack the selected spectra and measure the velocity dispersion of the resulting bin. We repeat this procedure 1000 times for each bin and the resulting uncertainty is taken as the 1$\sigma$ uncertainty of the measured velocity dispersions. To test the latter, we take the original stacked spectra and add random noise to each the spectra before fitting. The fitting is done on the ``noisy'' spectrum. We perform this procedure again 1000 times,  and the resulting uncertainty is taken again as the 1$\sigma$ dispersion of the obtained measurements. The total uncertainty due to statistical effects is the sum in quadrature of the two uncertainties. These are the error bars plotted in Figure~\ref{fig:fig4} and in the Appendix.

\section{Results and Discussion}
\label{results}

\begin{figure}[h]
\centering
\epsscale{1.0}\plotone{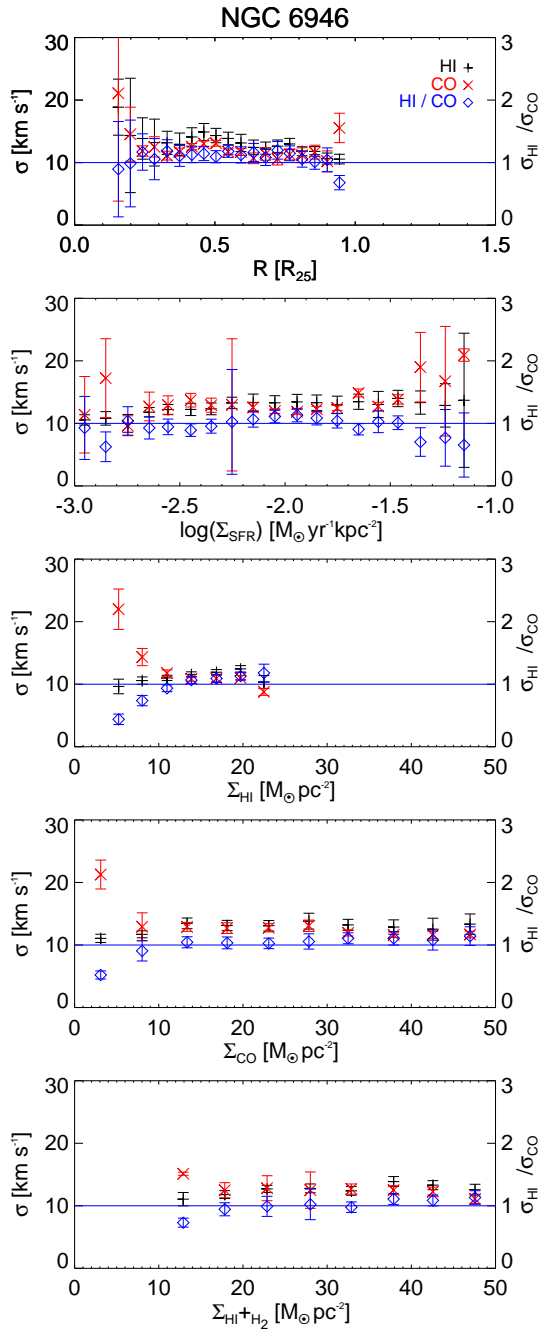}
\caption{Trends of velocity dispersion with different stacking parameters for  NGC 6946 (see Appendix for all other galaxies). From top to bottom we show results for stacking by galactocentric radius, and surface density of SFR, \hi, H$_{2}$, and total gas (different x-axes). The left y-axis gives the velocity dispersion, the right y-axis their ratio $\sigma_{\rm HI}/\sigma_{\rm CO}$. Black plus signs represent for \hi, red crosses those for CO, and blue diamonds represent the ratio of $\sigma_{\rm HI}/\sigma_{\rm CO}$. Error bars represent statistical uncertainties. The blue horizontal line indicates a value of unity for the ratio (right y-axis).}
\label{fig:fig4}
\end{figure}

\subsection{Trends of Velocity Dispersion with Stacking Parameters}
\label{vdtrends}
As a fundamental result from fitting the stacked spectra, we obtain measurements of \hi\ and CO velocity dispersions as a function of stacking parameter. These results are shown in Figure~\ref{fig:fig4} for NGC 6946. Equivalent plots for the remaining galaxies are shown in the Appendix.  

The top panel shows the variation of $\sigma_{\rm HI}$ and  $\sigma_{\rm CO}$ with galactocentric radius. Over large parts of the disks probed here ($R \lesssim R_{25}$) the velocity dispersion of both \hi\, and CO are roughly constant (ratio $\sim$\,1) and have typical values within $7-21$ \kmpers. The largest range in velocity dispersions are present near galaxy centers ($R \lesssim 0.2\ R_{25}$) where the largest velocity dispersion values are found, likely a signature of large velocity gradients inside our $\sim$500 pc beam caused for example by molecular bars. However, most of these points are excluded in our analysis due to double peaked spectral profiles (Section~\ref{data}). The error bars represent the statistical uncertainties in the velocity dispersion determination due to the number of stacked spectra and observational noise (Section~\ref{stat}). In the Appendix we show a table (Table \ref{tab5}) for each galaxy with the values of the \hi\, and \rm{CO} velocity dispersions as a function of galactocentric distance.

\begin{deluxetable}{c c c c c c} 
\tablecolumns{6}
\tablewidth{\columnwidth} 
\tablecaption{Spearman's Rank Correlation Coefficients for the complete sample of galaxies \label{corr}}
\tablehead{\colhead {} &\colhead{   R$_{25}$} &\colhead{$\Sigma_{\rm{SFR}}$} &\colhead{$\Sigma_{\rm{HI}}$} &\colhead{$\Sigma_{\rm{CO}}$} &\colhead{$\Sigma_{\rm{HI}}$+$\Sigma_{\rm{CO}}$}} 
\startdata
$\sigma_{\rm {HI}}$ & -0.6$\pm$0.5 & 0.9$\pm$0.3 & \phn0.6$\pm$0.6 & \phn0.3$\pm$0.5 & \phn0.6$\pm$0.5 \\
$\sigma_{\rm {CO}}$ & -0.5$\pm$0.4 & 0.5$\pm$0.6 & -0.8$\pm$0.4 & 0.2$\pm$0.6 & -0.2$\pm$0.6 \\
$\sigma_{\rm {HI}}$/$\sigma_{\rm{CO}}$ & \phn0.3$\pm$0.3 & 0.0$\pm$0.5 & \phn1.0$\pm$0.3 & \phn0.0$\pm$0.5 & \phn0.6$\pm$0.4
\enddata
\end{deluxetable}

The other panels in Figure~\ref{fig:fig4} show the variation of velocity dispersion when stacking by SFR surface density, \hi\, surface density, H$_2$ surface density, and total gas surface density, respectively. To first order, we do not observe a clear general trend of the velocity dispersions with any of these parameters. To quantify this assertion we calculate the Spearman's rank correlation coefficients for each galaxy and for each stacking parameter. That is, for each galaxy we calculate the coefficients for $\sigma_{\rm HI}$, $\sigma_{\rm CO}$, and $\sigma_{\rm HI}$/$\sigma_{\rm CO}$ for each of the five different stacking parameters: $R_{25}$, $\Sigma_{\rm SFR}$, $\Sigma_{\rm HI}$, $\Sigma_{\rm H2}$, $\Sigma_{\rm HI+H2}$. We present the median values of these coefficients in Table~\ref{corr}, together with their 1$\sigma$ range. Value of $\pm$1 indicate clear correlations, whereas values closer to 0 signify no correlation.

Naively, we would have expected a strong correlation between SFR surface density and velocity dispersion. This is because young massive stars are thought to be important contributors in driving turbulence and injecting energy into the ISM.  The galaxies in our sample show different behaviors: while for 7 galaxies the velocity dispersions are roughly constant with SFR surface density, for the remaining 5 (NGC 2903, NGC 3351, NGC 4736, NGC 5055, and NGC 6946) there is an increase of velocity dispersion with increasing SFR surface density. The increase in velocity dispersions go from $\sim\,30\%\,-\, \sim\,60\%$.  However, the true effect of SFR surface density on velocity dispersion remains inconclusive, as these two parameters vary most strongly with radius. Thus, there could be an additional radially varying effect that may explain the observed trends (e.g. shear due to differential rotation).

We find a similar result when stacking by \hi, H$_{2}$, and total gas surface densities. The measured velocity dispersions do not present a strong correlation with any of these stacking parameters, except again for NGC 5055.
 
\begin{figure*}
\centering
\epsscale{0.72}\plotone{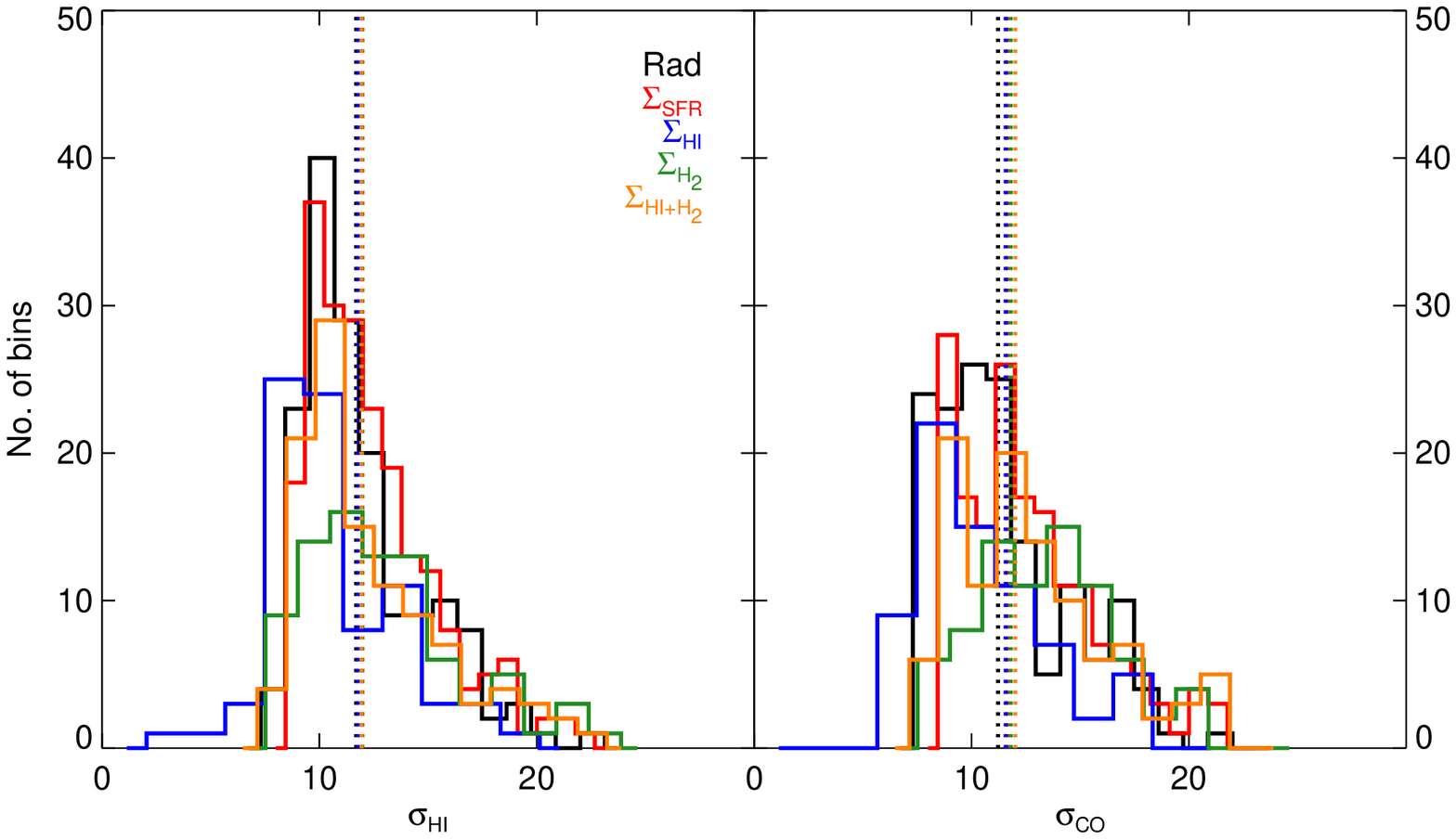}
\epsscale{0.38}\plotone{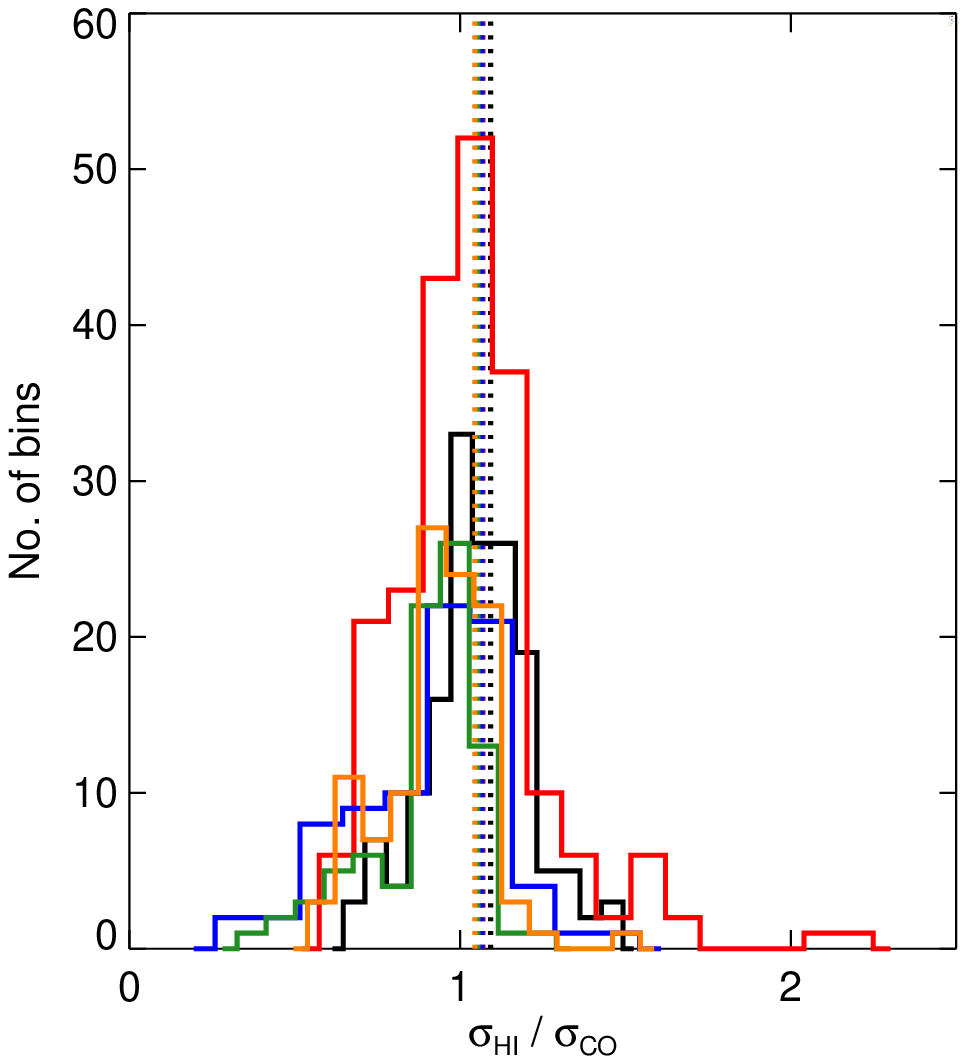}
\caption{Histograms of the $\sigma_{\rm HI}$ and $\sigma_{\rm CO}$ (left two panels) for each stacking parameter and their ratio $\sigma_{\rm HI} / \sigma_{\rm CO}$ (right panel). In black thin line are histogram of data stacked by galactocentric radius, in red thicker line stacked by SFR surface density, in blue thicker line for \hi\, surface density, in green thicker line for H$_{2}$ surface density, and in orange and thickest line for by total neutral gas surface density. The dashed lines show the median values obtained for the different stacking parameters (same color coding as with the histograms).}
\label{fig:fig5}
\end{figure*}

\begin{deluxetable}{r r r}
\tablecolumns{3}
 \tablewidth{0pt}
\tablecaption{Median values of the velocity dispersions measured in the whole sample of galaxies \label{tab3}}
 \tablehead{\colhead{\hi} & \colhead{\rm{CO}} & \colhead{\hi\,/\rm{CO}} \\
 \colhead{\kmpers} &  \colhead{\kmpers} & \colhead{ }}
 \startdata
 $11.9 \pm 3.1$ & $12.0 \pm 3.9$ & $1.0 \pm 0.2$
 \enddata
 \end{deluxetable} 
\vspace{2mm}
The trends for individual galaxies and for different stacking parameters are shown in Figure~\ref{fig:fig4} and the Appendix. Here we summarize the trends for our entire sample as histograms in left panels of Figure~\ref{fig:fig5} where we only include bins between 0.2$-$1 R$_{25}$ (Section \ref{galmod}). We find the \hi\ and CO velocity dispersions to have a range of $\sim 7-21$ \kmpers. The histograms for the ratio of \hi\ to CO velocity dispersions are shown in the right panel of Figure~\ref{fig:fig5}. Regardless of the stacking parameter we find the ratio to remain surprisingly constant at an average value of \ratmean. We summarize the median velocity dispersion values for our sample in Table~\ref{tab3}.

\subsection{Comparison to Previous Measurements \\ of Velocity Dispersions} 

A number of studies have been done in the past to constrain the \hi\ and CO velocity dispersions. The most straightforward comparison of our \hi\ velocity dispersions is the work done by \citet{ta09} as they analyze the same data (from THINGS). They find that the velocity dispersion decreases with  increasing radial distance from the galaxy center out to $\sim$4 R$_{25}$. Their method is to take the second moment value as a measure of the velocity dispersion regardless of the shape of the spectrum. We note that this can lead to an overestimate of the real velocity dispersions in regions were bulk motions are important, e.g., at stellar bars. Restricting their sample to the radial range covered by our analysis ($0.2 -1$ R$_{25}$), we find comparable velocity dispersions which are to first order constant (NGC 628, NGC 3184, NGC 3351, NGC 4736, NGC 5055, NGC 5194, and NGC 6946). Only for a subset of our sample (NGC 925, NGC 2903, NGC 4736, NGC 5055, and NGC 5194) we find a slight decrease of velocity dispersions with radius. It is important to note that we do not doubt that the velocity dispersions decrease at large galactocentric radii, e.g., the general conclusion of \citet{ta09}. We solely state that within the optical disk where both \hi\ and CO emission is strong we find no significant radial trends.

\renewcommand{\arraystretch}{1.2}
\begin{deluxetable*}{r r r r r r r}[t]
\tablecolumns{7}
\tablecaption{Comparison of CO Velocity Dispersions  \label{tab4}}
 \tablehead{
 \colhead{Galaxy} & \colhead{This work} & \colhead{Lit.} & \colhead{Lit. transition} & \colhead{ Beam size} & \colhead{$\Delta$R } & \colhead{Ref.} \\
 \colhead{ } & \colhead{km s$^{-1}$ }& \colhead{km s$^{-1}$} & \colhead{ } & \colhead{arcsec} & \colhead{ R$_{25}$} }
\startdata
NGC \phn 628  &  8.7\,$\pm$\,0.6  &  6.5\,$\pm$\,2.5 & \textbf{2\,--\,1}, 1\,--\,0 & 12, 23 & $ 0\phn \,-\,0.34 $ & \citet{co97} \\
\hline
NGC 5194 & 16.4\,$\pm$\,1.1 & 17.0\,$\pm$\,3.3 & 1\,--\,0 & 22 & $0.2\,-\,0.43$ & \citet{pe13} \\
\hline
NGC \phn 628 & 8.8\,$\pm$\,0.6  &  4.1\,$\pm$\,2.5 & \multirow{5}{*}{3\,--\,2} & \multirow{5}{*}{14.5} & $ 0\phn \,-\,0.5\phn $ & \multirow{5}{*}{\citet{wi11}} \\
NGC 2403 & 11.7\,$\pm$\,2.9  &  5.2\,$\pm$\,4.9 &  &  & $ 0\phn \,-\,0.45 $ &  \\
NGC 3184 & 9.8\,$\pm$\,1.8  &  6.8\,$\pm$\,5.9 &  &  & $ 0\phn \,-\,0.65 $ &  \\
NGC 4736 & 17.7\,$\pm$\,1.3  &  12.0\,$\pm$\,9.0 &  &  & $ 0\phn \,-\,0.3\phn $ &  \\
NGC 5055 & 15.2 \,$\pm$\,4.2  &  9.4\,$\pm$\,6.6 &  &  & $ 0\phn \,-\,0.55 $ & 
\enddata
\end{deluxetable*}

Only a few measurements of CO velocity dispersions in other galaxies have been published so far. We summarize these together with our measurements in Table \ref{tab4}. \citet{wi11} carried out a study of the \cothree\ line in 12 nearby galaxies with the JCMT single dish telescope. They measure values for the CO velocity dispersion in the range of $4-20$ \kmpers. For the five galaxies that overlap with our sample we list the average velocity dispersion measurements in Table \ref{tab4}. While both studies find no clear signature of a radial trend, the velocity dispersion values by \citet{wi11} are in general $\sim$40\% smaller than our measurements. It is not immediately clear if this (apparent) difference relates to a true difference in velocity dispersions for \cotwo\ and \cothree\ gas disks or if it relates to the limited signal-to-noise in the JCMT data. \citet{co97} studied the two nearly face--on galaxies NGC\,628 and NGC\,3938 and found similar \hi\ and CO vertical velocity dispersions in selected lines--of--sight as we do in our study.

Most recently, the molecular gas of M51 was mapped with high sensitivity by the PAWS survey \citep{sh13} and velocity dispersions were studied by \citet{pe13}. They compared CO single dish measurements from the IRAM-30 m telescope with interferometric observations taken with PdBI. The velocity dispersions they measure with interferometric-only data are $\sim$50\% lower than the ones measured with the single-dish telescope. Their single dish measurements are comparable to the results we obtain. They interpret this difference in velocity dispersions as evidence of a high-dispersion diffuse CO component, in addition to the clumpy molecular disk. 

\subsection{Multi--component Phase Structure}

As stated in Section~\ref{intro}, the ISM does not have sharp boundaries between the different phases. There is a natural transition from atomic to molecular gas, where the highest densities and lowest temperatures are reached. The two-phase model for the neutral atomic gas \citep{wo03,fi69} differentiates between the WNM (broad component) and CNM (narrow component), both contributing to the total \hi\ emission. However, there is also an unstable phase which may account for up to 50\% of the \hi\ mass. In a simple picture, the thermal broadening of the warm neutral \hi\ medium at a temperature of 8000 K leads to a velocity dispersion of $\sim$8 \kmpers. We measure values for $\sigma_{\rm HI}$ in the range of $7-21$ \kmpers, with a median value of \hivmed\ and a mean of \hivmean\ with a 1$\sigma$ dispersion of \hidisp. Within the uncertainties this is consistent with the presence of {\em only} an  `undisturbed' WNM, but there is likely additional broadening present due to non--thermal gas motions.

On the contrary, in the usually accepted picture one would expect much smaller velocity dispersions for the cold molecular gas phase. Here, the clouds are thought to be few 10\,pc in size and have observed velocity dispersions of 2$-$5 \kmpers\ \citep{ma07,bo08}. The velocity dispersions measured by us exceed this value by a wide margin: we measure values for $\sigma_{\rm CO}$ ranging within $6.7-23$ \kmpers\, with an average mean value of \covmean\ with a 1$\sigma$ dispersion of \counc, very close to what we measure for the \hi. These values are much larger than the values expected by purely thermal processes ($\sigma \sim 0.1$ \kmpers\, for 10~K gas) and are also larger than those expected by intra--cloud motions, which in M\,31 are $\sim 6$ \kmpers\ (Schruba et al., in prep.). At the resolution we are working, we cannot reject the possibility that part of the dispersion we are measuring could be due to bulk motions unresolved by the velocity maps at ~ 11 arcsec (the native natural-weighted THINGS resolution). However, this would affect both \hi\, and CO measurements.  

Our main result is a rough agreement of the velocity dispersions for both \hi\ and CO components. This is in agreement with \citet{co97} that studied NGC 628 and NGC 3938. Based on their measurements they argue that: ``The similarity of the CO and \hi\ dispersions suggests that the two components are well mixed, and are only two different phases of the same kinematical gas component''---an interpretation that could also apply to our measurements.

For galaxies with large scale disk morphology the observed (macroscopic) velocity dispersions can be related to the vertical density profile of the observed tracer. The density profile is set by hydrostatic equilibrium such that gravitational weight is balanced by gas pressure, i.e., its velocity dispersion. For the general case of a multi-component stellar and gaseous disk the vertical density profile of each component needs to balance the combined gravitational field of stars, gas, and dark matter. Useful approximations to the full numerical solution are given by \citet{el89,na02}. In general, an increase of velocity dispersion causes an increase in scale height at fixed gravitational field (i.e., surface density) while an increase in surface density decreases the scale height at fixed velocity dispersion. The hydrostatic equilibrium assumption also let us link the observables for face--on galaxies (i.e., velocity dispersion) and edge--on galaxies (i.e., scale height).

Evidence for a thick molecular gas disk in nearby galaxies has been presented in several studies. As highlighted in the introduction, early observations of the edge--on galaxy NGC\,891 revealed the presence of an extended CO component \citep{ga92}, which has in the meanwhile been confirmed by deeper and more extended maps of the galaxy (Garcia-Burillo et al., priv.\ comm.). A massive extraplanar reservoir of molecular gas has also been reported in the highly inclined galaxy M\,82 \citep{wa02}, though this material has likely been expelled from the disk by the intense star formation activity and thus may not be in hydrostatic balance.

A complementary method to infer the vertical disk structure has been performed by \citet{co12}. They analyzed the power spectra of gas and infrared emission in M\,33 and observed a break in the power spectrum tracing the transition from 2D to 3D turbulence suggestively tracing the disk scale height. They detect a thick disk for both HI and CO of comparable height.

Nevertheless observational evidence is scarce to date, and in some cases circumstantial, as it is impossible to directly measure the vertical dispersion of the CO gas (requiring face--on orientation) and the thickness of a CO disk (edge--on orientation) in the same system.

Observationally disentangling the thickness of the atomic and molecular gas disk in our own Galaxy is difficult given our in--disk viewing position and severe distance ambiguities. Furthermore, prominent structures are known that extend far away from the galactic plane such as the Goult's Belt and the Taurus--Perseus--Auriga association. Employing large--scale CO mapping, \citet{da94} found evidence for a homogeneous thick molecular disk $\sim 3$ times more extended than the (well known) thin disk and comprising at least 5\% of the entire CO luminosity. However, as mm single-dish observations need to perform `differential' measurements to subtract atmospheric emission, they are largely insensitive to any extended emission and leave the true mass of a diffuse gas component large unconstrained. In addition, a number of high--latitude clouds are known in the literature \citep[e.g.,][]{ma85, ma96} though at least some of them are from extragalactic origin and thus will not be in hydrostatic balance.

Estimates for external galaxies go as far as they may contain a diffuse medium that accounts for $\sim 30\%-50\%$ of the molecular mass as suggested by \citep{ga92, pe13}, however, additional observations would be required to assert a general validity.

\subsection{Implications}

With the current data it is hard to estimate how much of the observed luminosity emerges from a thin versus a thick disk component. However, it is clear that the contribution (in flux) from the thick component must be significant, of order $\sim$30\%, in order to be detectable. We can only speculate what fraction of the molecular mass would be entrained in such a thick component: the conditions of the gas (gas temperature, radiation field, and resulting gas excitation) will likely be significantly different compared to the GMC environment in the star-forming disk. This has two very direct implications. The first one is the importance of quantifying how much the CO-to-H$_2$ conversion factor ($X_{\rm CO}$) varies between dense and diffuse molecular gas. It is not clear yet whether these different environments would necessarily lead to a difference in the conversion factor. A recent study of the high--latitude cloud MBM\,40 gave a conversion factor for this cloud which is consistent with the Galactic factor \citep{co13}. \citet{da01} and \citet{li10} find something similar, as they conclude that $X_{\rm CO}$ shows little variation with latitude from the galactic value, however, they note that further studies should be conducted to reach a more definite answer on this issue. The second is that if a significant fraction (of order 30\%) of the CO line flux was indeed due to a component that is not coincident with the star-forming disk, then previous studies would have overestimated the flux that is co--spatial with ongoing star formation. This would imply that the derived molecular gas masses would have to be decreased by a similar factor. \citet{she13} investigated the SFR\,--\,CO relation in the STING sample finding indications of a sub-linearity for 8 them. One possible interpretation they give is that there might be ``excited CO in the diffuse or non-star-forming ISM''. Future observations using both interferometers and single dish telescopes are needed to clearly separate the relative flux contribution of a diffuse versus a compact molecular gas phase in nearby galaxies.

It is also noteworthy that CO appears to be able to survive in environments that are significantly different than those found near the star-forming disk, i.e., at significantly lower extinction and resulting shielding. The existence of a substantial thick molecular disk also challenges various models of H$_2$ formation. This includes models that rely on Parker instability (i.e., colliding flows and flows along magnetic field lines), as in those models dense gas settles near the galactic mid-plane forming a thin disk. On the other hand, models assuming gravitational collapse as means of dense molecular gas formation do not face these issues \citep{ma07}.

\section{Summary}
\label{sum}

We present resolved measurements of \hi\ and CO velocity dispersions in 12 nearby galaxies taken from the THINGS and HERACLES surveys. To construct spectra of high significance, we stack spectra within different bins defined by physical parameters: galactocentric distance, star formation rate surface density, \hi\ surface density, H$_{2}$ surface density, and total gas surface density. The velocity dispersions are determined by fitting a single Gaussian profile to the stacked spectra.

For each galaxy we obtain a range of measurements of the \hi\ and CO velocity dispersions and their ratio as function of each stacking parameter. We do not find a strong trend of the velocity dispersions with any of the stacking parameters. We find a mean value $\sigma_{\rm HI}$ of \hisigmean\ and a mean value of $\sigma_{\rm CO}$ of \cosigmean, their ratio is on average $\sigma_{\rm HI}/\sigma_{\rm CO} =$ \ratmean. The latter value is surprising, as the expectation would have been that the CO dispersion is significantly smaller than that of the \hi. This strongly points towards the existence of a diffuse and `thick' molecular phase with scale heights comparable to that of the \hi. Evidence for such thick molecular gas disks has been presented in earlier studies for individual systems, and our results show that they may be a common phenomenon among galaxies. If indeed a significant fraction of the total CO flux was due to a component that is not coincident with the star-forming disk, then this needs to be taken into account, e.g., when deriving masses of the molecular gas component that is associated with star formation in a galaxy's disk.

Indisputable evidence for the existence of a thick molecular gas disk can only come from observations of edge--on galaxies, such as the work on NGC\,891 \citep[][and Garcia--Burillo, priv.\ comm.]{ga92} or M\,82 \citep{wa02}, so it is critical to obtain more such measurements in nearby close--to--edge--on galaxies (e.g. NGC\,4013, NGC\,4244, NGC\,4565, NGC\,5866, among others). An alternative promising way is to follow the procedure by \citet{pe13} to compare interferometric and single--dish imaging of nearby close--to--face--on galaxies (that are more numerous than edge--on systems). In the former interferometric observations, any extended/diffuse component will be invisible, thus only revealing the presence of compact GMC--like structures. In the latter single--dish observations {\em all} the emission, including a potential thick and high--dispersion molecular gas disk, will be recorded, albeit at lower resolution. A comparison of the two should thus give further insights on the frequency of thick molecular gas disks in spiral galaxies.

\acknowledgements

We thank the anonymous referee for very favorable and thoughtful comments. ACP acknowledges support from the DFG priority program 1573 `The physics of the interstellar medium' , from the IMPRS for  Astronomy \& Cosmic Physics at the University of Heidelberg, and from the Mexican National Council for Science and Technology (CONACYT). AS and FB thank Mark Krumholz for stimulating discussion on the implications for models of H$_2$ formation.

\bibliography{mybib}

\appendix 

Here we present the measurements of \hi\ (black plus signs) and CO (red crosses) velocity dispersions as a function of each of the stacking parameters for all galaxies in our sample. For each galaxy, the uppermost plot corresponds to the measurements we obtain from stacking by radial distance to the center of the galaxy. The second plot down is the equivalent for SFR surface density stacking. The third corresponds to \hi\ surface density stacking, the fourth plot to H$_{2}$ surface density stacking, and finally the last plot corresponds to total neutral gas surface density stacking. On the same plots we also present the values of the ratio of $\sigma_{\rm HI}/\sigma_{\rm CO}$ in blue diamonds (right y-axis label).

\begin{figure}[b]
	\epsscale{1}\plotone{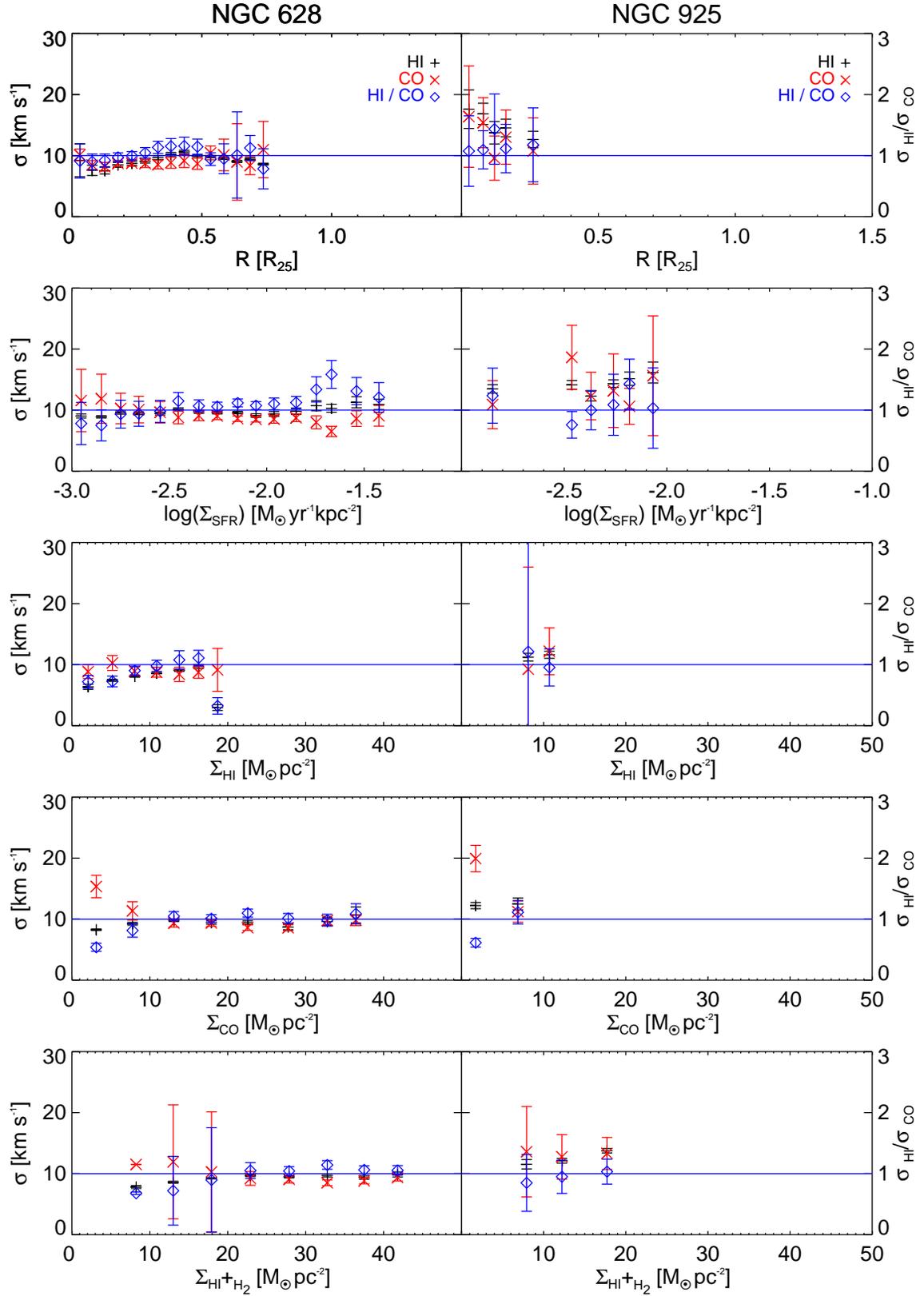}
	\caption{Trends of velocity dispersion with different stacking parameters. From top to bottom we present the results for galactocentric radius, SFR density, HI surface density, H$_{2}$ surface density, and total neutral gas surface density (different x-axes). On the x-axis we show the value of the corresponding stacking parameter. On the left y-axis we show the value of the velocity dispersion, and on the right y-axis is the value of the $\sigma_{\rm HI}/\sigma_{\rm CO}$ ratio. Black plus signs represent the values for \hi\ velocity dispersion, red crosses correspond to the CO values, and blue diamonds represent the ratio of $\sigma_{\rm HI}/\sigma_{\rm CO}$. The error bars represent statistical uncertainties as discussed in Section~\ref{stat} . The blue horizontal line indicates a value of unity for the ratio (right y-axis).\label{fig:fig7}}
\end{figure}

\begin{figure} 
	\epsscale{1}\plotone{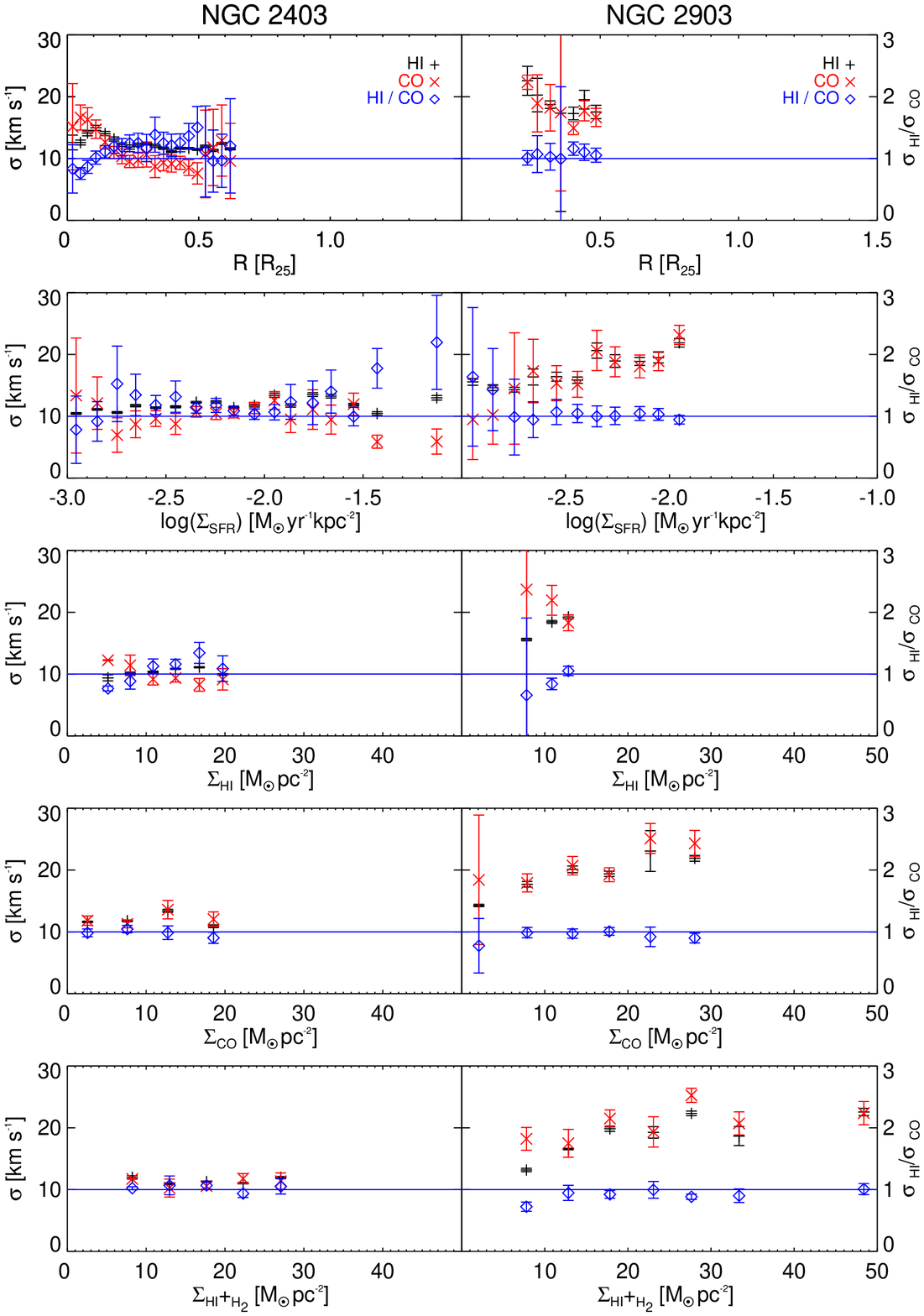}
	\caption{(cont.)}
\end{figure}

\begin{figure}
	\epsscale{1}\plotone{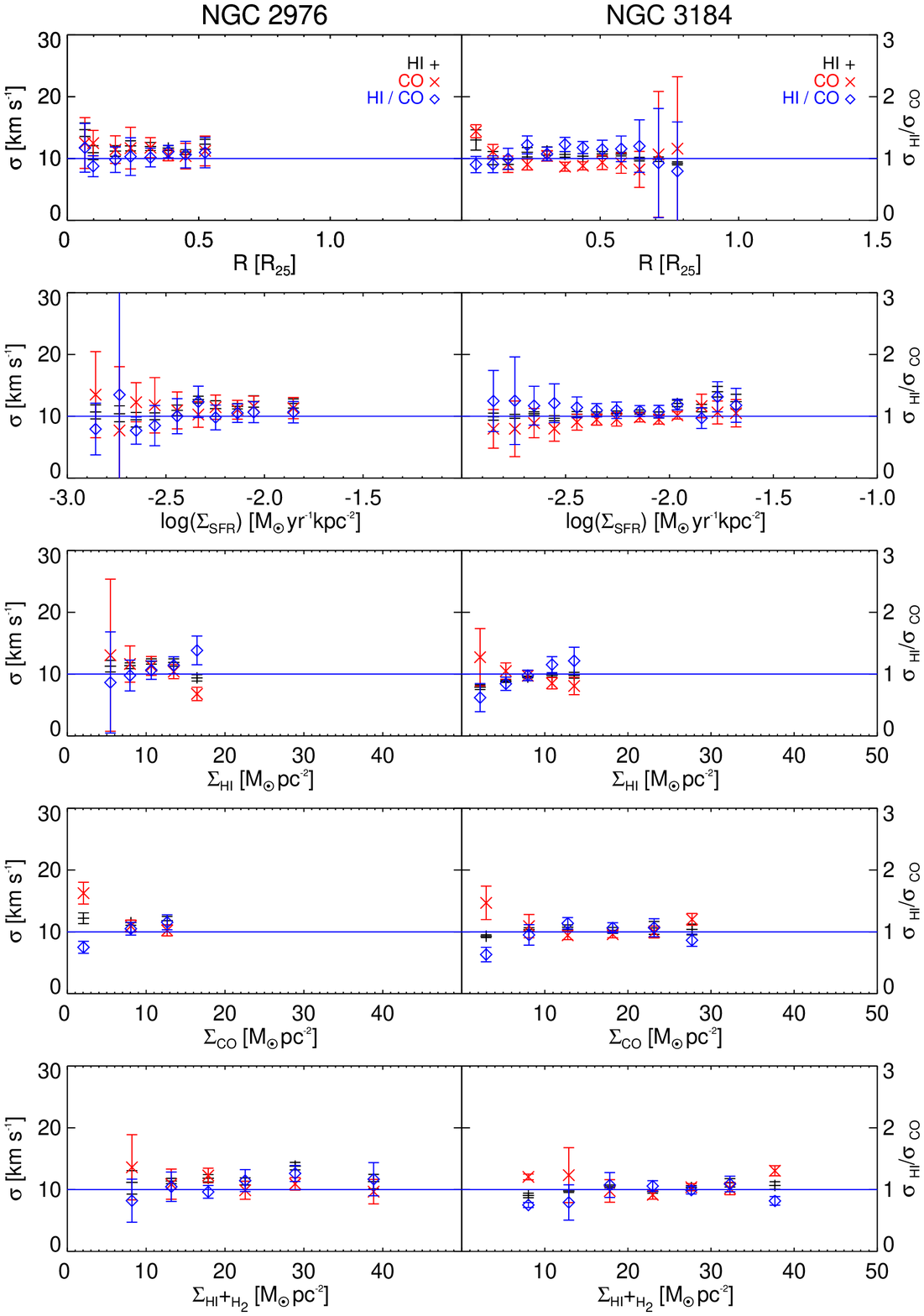}
	\caption{(cont.)}
\end{figure}

\begin{figure}
	\epsscale{1}\plotone{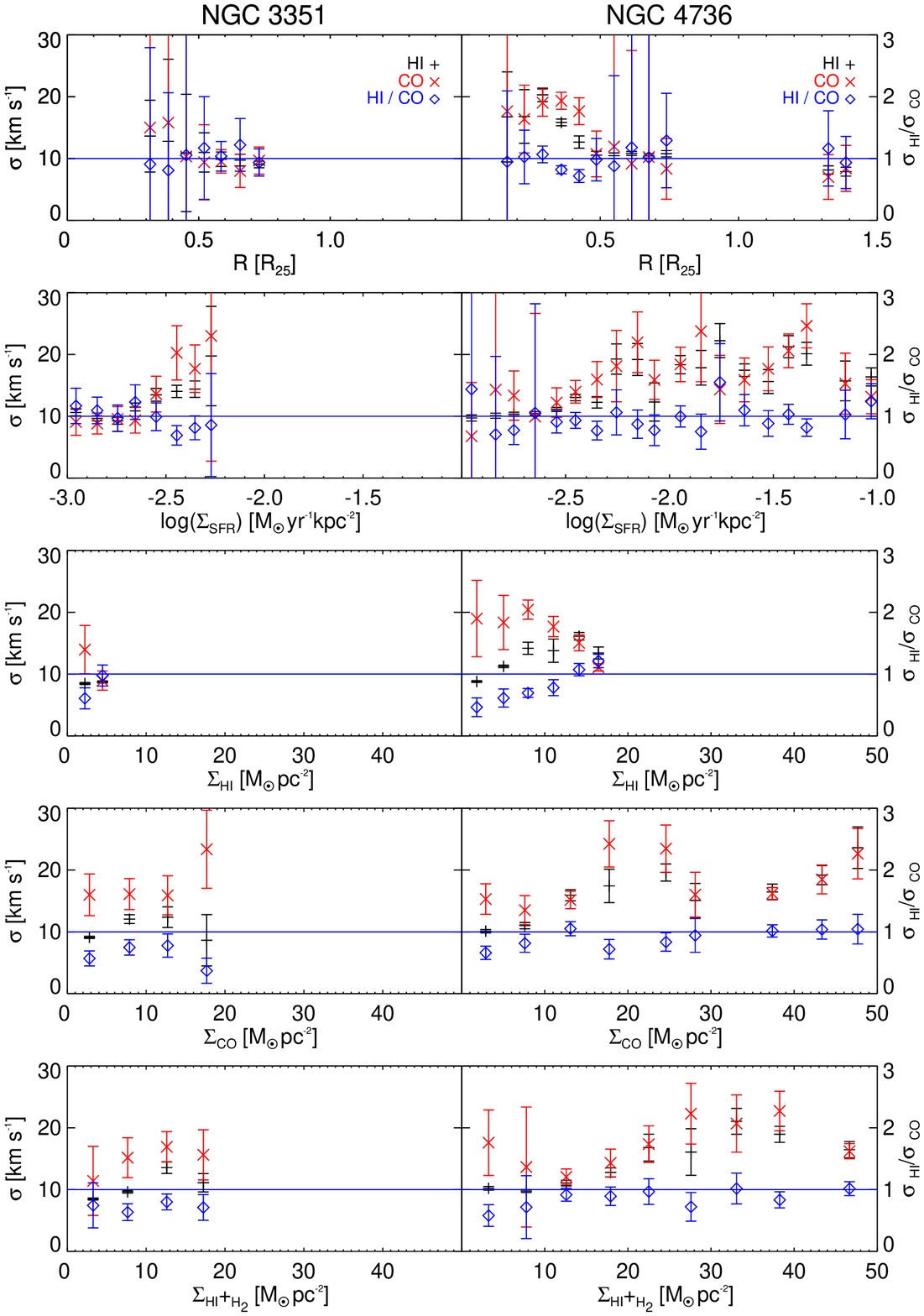}
	\caption{(cont.)}
\end{figure}

\begin{figure}
	\epsscale{1}\plotone{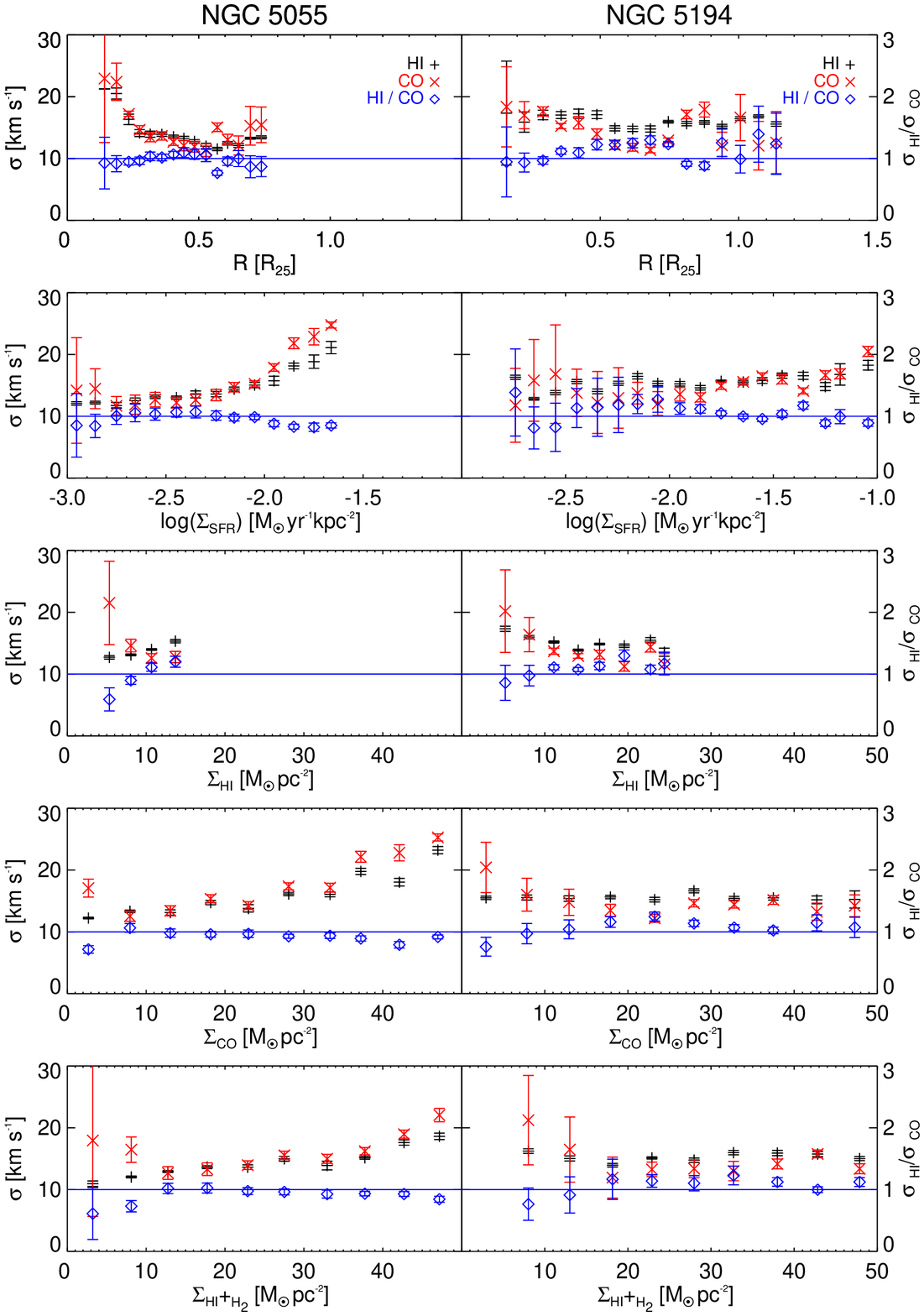}
	\caption{(cont.)}
\end{figure}

\begin{figure}
	\epsscale{0.5}\plotone{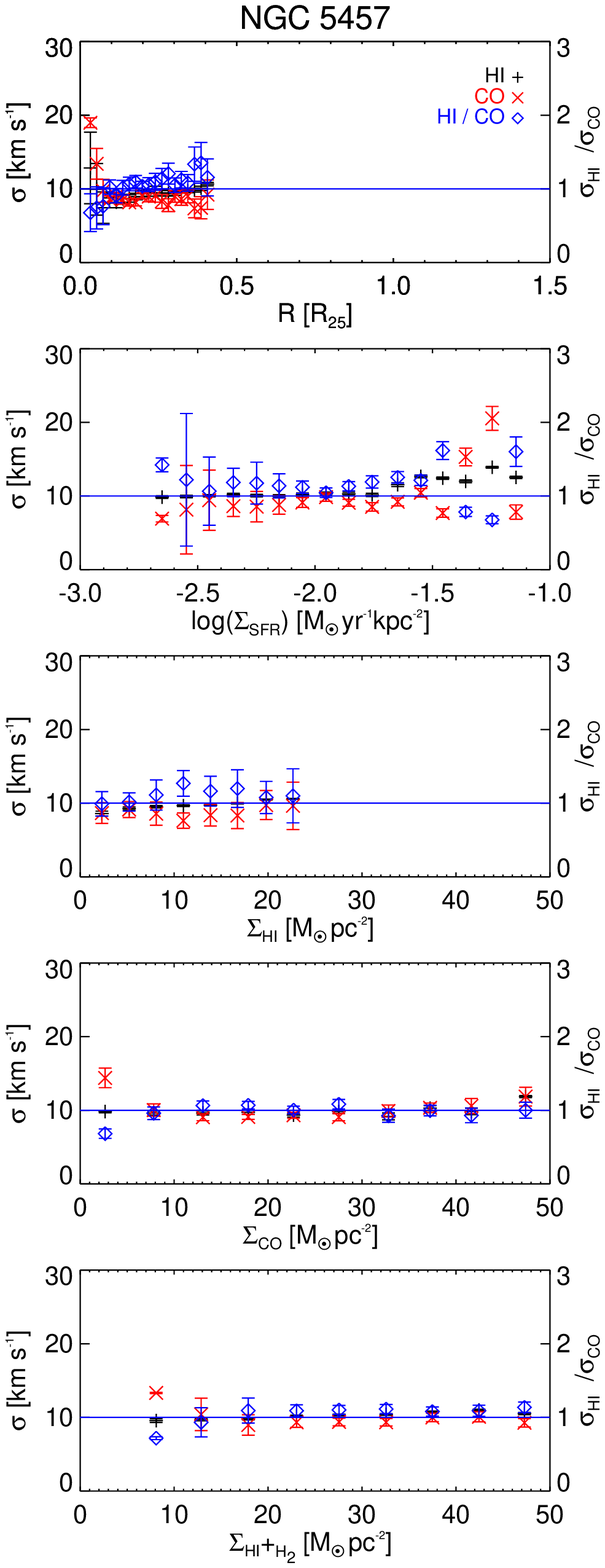}
	\caption{(cont.)}
\end{figure}

\clearpage

\renewcommand{\arraystretch}{1.2}
\begin{deluxetable}{c r r r c}
\tablecolumns{5}
\tablewidth{0pt}
\tablecaption{Velocity Dispersions Measured in Radial Bins \label{tab5}}
\tablehead{\colhead{ Galaxy  } & \colhead{ R } & \colhead{  $\sigma_{\rm{HI}}$  } & \colhead{  $\sigma_{\rm{CO}}$  } & \colhead{  $\sigma_{\rm{HI}}$/$\sigma_{\rm{CO}}$  } \\\colhead{ } & \colhead{R$_{25}$} & \colhead{km\,s$^{-1}$} & \colhead{km\,s$^{-1}$} & \colhead{ }}
\startdata
\multirow{4}{*}{NGC 628} & 0.03 & 9.2\,$\pm$\, 2.7 & 10.1\,$\pm$\, 1.0 & 0.9\,$\pm$\, 0.3 \\
 & 0.08 & 7.6\,$\pm$\, 0.8 & 8.4\,$\pm$\, 0.8 & 0.9\,$\pm$\, 0.1 \\
 & 0.13 & 7.5\,$\pm$\, 0.5 & 8.2\,$\pm$\, 0.7 & 0.9\,$\pm$\, 0.1 \\
 & \vdots \hspace{2mm} & \vdots \hspace{6mm} & \vdots \hspace{6mm} & \vdots \\ \hline
\multirow{4}{*}{NGC 925} & 0.03 & 17.6\,$\pm$\, 3.2 & 16.4\,$\pm$\, 8.3 & 1.1\,$\pm$\, 0.6 \\
 & 0.08 & 16.8\,$\pm$\, 1.7 & 15.4\,$\pm$\, 4.1 & 1.1\,$\pm$\, 0.3 \\
 & 0.12 & 13.7\,$\pm$\, 1.8 & 9.6\,$\pm$\, 3.6 & 1.4\,$\pm$\, 0.6 \\
 & \vdots \hspace{2mm} & \vdots \hspace{6mm} & \vdots \hspace{6mm} & \vdots \\ \hline
\multirow{4}{*}{NGC 2403} & 0.02 & 12.3\,$\pm$\, 1.2 & 15.3\,$\pm$\, 7.0 & 0.8\,$\pm$\, 0.4 \\
 & 0.05 & 12.0\,$\pm$\, 0.4 & 16.4\,$\pm$\, 2.0 & 0.7\,$\pm$\, 0.1 \\
 & 0.08 & 12.4\,$\pm$\, 0.4 & 14.2\,$\pm$\, 1.9 & 0.9\,$\pm$\, 0.1 \\
 & \vdots \hspace{2mm} & \vdots \hspace{6mm} & \vdots \hspace{6mm} & \vdots \\ \hline
\multirow{4}{*}{NGC 2903} & 0.24 & 20.8\,$\pm$\, 2.3 & 18.7\,$\pm$\, 1.2 & 1.1\,$\pm$\, 0.1 \\
 & 0.27 & 19.2\,$\pm$\, 2.7 & 17.4\,$\pm$\, 4.6 & 1.1\,$\pm$\, 0.3 \\
 & 0.32 & 17.3\,$\pm$\, 0.6 & 15.7\,$\pm$\, 3.8 & 1.1\,$\pm$\, 0.3 \\
 & \vdots \hspace{2mm} & \vdots \hspace{6mm} & \vdots \hspace{6mm} & \vdots \\ \hline
\multirow{4}{*}{NGC 2976} & 0.07 & 14.7\,$\pm$\, 1.1 & 12.2\,$\pm$\, 4.1 & 1.2\,$\pm$\, 0.4 \\
 & 0.10 & 10.5\,$\pm$\, 1.1 & 12.3\,$\pm$\, 2.1 & 0.9\,$\pm$\, 0.2 \\
 & 0.18 & 10.8\,$\pm$\, 0.9 & 10.5\,$\pm$\, 2.2 & 1.0\,$\pm$\, 0.2 \\
 & \vdots \hspace{2mm} & \vdots \hspace{6mm} & \vdots \hspace{6mm} & \vdots \\ \hline
\multirow{4}{*}{NGC 3184} & 0.05 & 13.0\,$\pm$\, 1.6 & 14.4\,$\pm$\, 1.0 & 0.9\,$\pm$\, 0.1 \\
 & 0.11 & 10.1\,$\pm$\, 1.1 & 11.1\,$\pm$\, 1.2 & 0.9\,$\pm$\, 0.1 \\
 & 0.17 & 9.0\,$\pm$\, 0.8 & 9.1\,$\pm$\, 1.3 & 1.0\,$\pm$\, 0.2 \\
 & \vdots \hspace{2mm} & \vdots \hspace{6mm} & \vdots \hspace{6mm} & \vdots \\ \hline
\multirow{4}{*}{NGC 3351} & 0.31 & 13.6\,$\pm$\, 5.8 & 15.0\,$\pm$\,30.5 & 0.9\,$\pm$\, 1.9 \\
 & 0.38 & 12.8\,$\pm$\,13.3 & 15.8\,$\pm$\,18.3 & 0.8\,$\pm$\, 1.3 \\
 & 0.45 & 10.9\,$\pm$\, 9.5 & 10.3\,$\pm$\,66.0 & 1.1\,$\pm$\, 6.8 \\
 & \vdots \hspace{2mm} & \vdots \hspace{6mm} & \vdots \hspace{6mm} & \vdots \\ \hline
\multirow{4}{*}{NGC 4736} & 0.16 & 16.7\,$\pm$\, 7.3 & 17.7\,$\pm$\,19.9 & 0.9\,$\pm$\, 1.1 \\
 & 0.23 & 16.8\,$\pm$\, 4.3 & 16.4\,$\pm$\, 5.5 & 1.0\,$\pm$\, 0.4 \\
 & 0.29 & 20.3\,$\pm$\, 1.1 & 19.0\,$\pm$\, 2.2 & 1.1\,$\pm$\, 0.1 \\
 & \vdots \hspace{2mm} & \vdots \hspace{6mm} & \vdots \hspace{6mm} & \vdots \\ \hline
\multirow{4}{*}{NGC 5055} & 0.11 & 19.1\,$\pm$\, 0.0 & 23.1\,$\pm$\,10.4 & 0.8\,$\pm$\, 0.4 \\
 & 0.14 & 17.1\,$\pm$\, 0.9 & 18.9\,$\pm$\, 3.0 & 0.9\,$\pm$\, 0.2 \\
 & 0.19 & 17.0\,$\pm$\, 0.9 & 19.1\,$\pm$\, 0.4 & 0.9\,$\pm$\, 0.0 \\
 & \vdots \hspace{2mm} & \vdots \hspace{6mm} & \vdots \hspace{6mm} & \vdots \\ \hline
\multirow{4}{*}{NGC 5194} & 0.16 & 17.4\,$\pm$\, 8.4 & 18.4\,$\pm$\, 6.5 & 0.9\,$\pm$\, 0.6 \\
 & 0.23 & 15.9\,$\pm$\, 1.6 & 17.0\,$\pm$\, 2.2 & 0.9\,$\pm$\, 0.2 \\
 & 0.29 & 17.0\,$\pm$\, 0.8 & 17.6\,$\pm$\, 0.7 & 1.0\,$\pm$\, 0.1 \\
 & \vdots \hspace{2mm} & \vdots \hspace{6mm} & \vdots \hspace{6mm} & \vdots \\ \hline
\multirow{4}{*}{NGC 5457} & 0.03 & 12.8\,$\pm$\, 4.9 & 19.0\,$\pm$\, 0.7 & 0.7\,$\pm$\, 0.3 \\
 & 0.05 & 9.9\,$\pm$\, 3.5 & 13.4\,$\pm$\, 2.1 & 0.7\,$\pm$\, 0.3 \\
 & 0.07 & 7.4\,$\pm$\, 2.1 & 9.7\,$\pm$\, 1.6 & 0.8\,$\pm$\, 0.3 \\
 & \vdots \hspace{2mm} & \vdots \hspace{6mm} & \vdots \hspace{6mm} & \vdots \\ \hline
\multirow{4}{*}{NGC 6946} & 0.16 & 18.9\,$\pm$\, 4.5 & 21.1\,$\pm$\,17.2 & 0.9\,$\pm$\, 0.8 \\
 & 0.20 & 14.3\,$\pm$\, 9.1 & 14.5\,$\pm$\, 4.3 & 1.0\,$\pm$\, 0.7 \\
 & 0.24 & 13.9\,$\pm$\, 3.3 & 11.9\,$\pm$\, 0.8 & 1.2\,$\pm$\, 0.3 \\
 & \vdots \hspace{2mm} & \vdots \hspace{6mm} & \vdots \hspace{6mm} & \vdots
\enddata
\vspace{0.2cm}
\tablecomments{Here we present the first three radial bins for each galaxy. The rest of the bins are shown in the online version.}
\end{deluxetable}
\clearpage

\end{document}